
\message
{JNL.TEX version 0.92 as of 6/9/87.  Report bugs and problems to Doug Eardley.}
\message
{This is X.G. Wen's copy}

\catcode`@=11
\expandafter\ifx\csname inp@t\endcsname\relax\let\inp@t=\input
\def\input#1 {\expandafter\ifx\csname #1IsLoaded\endcsname\relax
\inp@t#1%
\expandafter\def\csname #1IsLoaded\endcsname{(#1 was previously loaded)}
\else\message{\csname #1IsLoaded\endcsname}\fi}\fi
\catcode`@=12



\font\twelverm=amr10 scaled 1200    \font\twelvei=ammi10 scaled 1200
\font\twelvesy=amsy10 scaled 1200   \font\twelveex=amex10 scaled 1200
\font\twelvebf=ambx10 scaled 1200   \font\twelvesl=amsl10 scaled 1200
\font\twelvett=amtt10 scaled 1200   \font\twelveit=amti10 scaled 1200
\font\twelvesc=amcsc10 scaled 1200  \font\twelvesf=amssmc10 scaled 1200
\skewchar\twelvei='177   \skewchar\twelvesy='60


\def\twelvepoint{\normalbaselineskip=12.4pt plus 0.1pt minus 0.1pt
  \abovedisplayskip 12.4pt plus 3pt minus 9pt
  \belowdisplayskip 12.4pt plus 3pt minus 9pt
  \abovedisplayshortskip 0pt plus 3pt
  \belowdisplayshortskip 7.2pt plus 3pt minus 4pt
  \smallskipamount=3.6pt plus1.2pt minus1.2pt
  \medskipamount=7.2pt plus2.4pt minus2.4pt
  \bigskipamount=14.4pt plus4.8pt minus4.8pt
  \def\rm{\fam0\twelverm}          \def\it{\fam\itfam\twelveit}%
  \def\sl{\fam\slfam\twelvesl}     \def\bf{\fam\bffam\twelvebf}%
  \def\mit{\fam 1}                 \def\cal{\fam 2}%
  \def\sc{\twelvesc}		   \def\tt{\twelvett}
  \def\sf{\twelvesf}
  \textfont0=\twelverm   \scriptfont0=\tenrm   \scriptscriptfont0=\sevenrm
  \textfont1=\twelvei    \scriptfont1=\teni    \scriptscriptfont1=\seveni
  \textfont2=\twelvesy   \scriptfont2=\tensy   \scriptscriptfont2=\sevensy
  \textfont3=\twelveex   \scriptfont3=\twelveex  \scriptscriptfont3=\twelveex
  \textfont\itfam=\twelveit
  \textfont\slfam=\twelvesl
  \textfont\bffam=\twelvebf \scriptfont\bffam=\tenbf
  \scriptscriptfont\bffam=\sevenbf
  \normalbaselines\rm}



\def\beginlinemode{\endmode
  \begingroup\parskip=0pt \obeylines\def\\{\par}\def\endmode{\par\endgroup}}
\def\beginparmode{\endmode
  \begingroup \def\endmode{\par\endgroup}}
\let\endmode=\par
{\obeylines\gdef\
{}}
\def\singlespace{\baselineskip=\normalbaselineskip}

\def\oneandahalfspace{\baselineskip=\normalbaselineskip
  \multiply\baselineskip by 3 \divide\baselineskip by 2}
\def\doublespace{\baselineskip=\normalbaselineskip \multiply\baselineskip by 2}

\newcount\firstpageno
\firstpageno=2
\footline={\ifnum\pageno<\firstpageno{\hfil}\else{\hfil\twelverm\folio\hfil}\fi}
\def\toppageno{\global\footline={\hfil}\global\headline
  ={\ifnum\pageno<\firstpageno{\hfil}\else{\hfil\twelverm\folio\hfil}\fi}}
\let\rawfootnote=\footnote		
\def\footnote#1#2{{\rm\singlespace\parindent=0pt\parskip=0pt
  \rawfootnote{#1}{#2\hfill\vrule height 0pt depth 6pt width 0pt}}}
\def\raggedcenter{\leftskip=4em plus 12em \rightskip=\leftskip
  \parindent=0pt \parfillskip=0pt \spaceskip=.3333em \xspaceskip=.5em
  \pretolerance=9999 \tolerance=9999
  \hyphenpenalty=9999 \exhyphenpenalty=9999 }
\def\dateline{\rightline{\ifcase\month\or
  January\or February\or March\or April\or May\or June\or
  July\or August\or September\or October\or November\or December\fi
  \space\number\year}}
\def\received{\vskip 3pt plus 0.2fill
 \centerline{\sl (Received\space\ifcase\month\or
  January\or February\or March\or April\or May\or June\or
  July\or August\or September\or October\or November\or December\fi
  \qquad, \number\year)}}


\hsize=6.5truein
\hoffset=0truein
\vsize=8.9truein
\voffset=0truein
\parskip=\medskipamount
\def\\{\cr}
\twelvepoint		
\doublespace		
\overfullrule=0pt	




\def\title			
  {\null\vskip 3pt plus 0.2fill
   \beginlinemode \doublespace \raggedcenter \bf}

\def\author			
  {\vskip 3pt plus 0.2fill \beginlinemode
   \singlespace \raggedcenter\sc}

\def\affil			
  {\vskip 3pt plus 0.1fill \beginlinemode
   \oneandahalfspace \raggedcenter \sl}

\def\abstract			
  {\vskip 3pt plus 0.3fill \beginparmode
   \oneandahalfspace ABSTRACT: }

\def\endtitlepage		
  {\endpage			
   \body}

\def\body			
  {\beginparmode}		

\def\head#1{			
  \goodbreak\vskip 0.5truein	
  {\immediate\write16{#1}
   \raggedcenter \uppercase{#1}\par}
   \nobreak\vskip 0.25truein\nobreak}

\def\beginitems{
\par\medskip\bgroup\def\i##1 {\item{##1}}\def\ii##1 {\itemitem{##1}}
\leftskip=36pt\parskip=0pt}
\def\enditems{\par\egroup}

\def\beneathrel#1\under#2{\mathrel{\mathop{#2}\limits_{#1}}}

\def\refto#1{$^{#1}$}		

\def\references			
  {\head{References}		
   \beginparmode
   \frenchspacing \parindent=0pt \leftskip=1truecm
   \parskip=8pt plus 3pt \everypar{\hangindent=\parindent}}

\def\referencesnohead   	
  {                     	
   \beginparmode
   \frenchspacing \parindent=0pt \leftskip=1truecm
   \parskip=8pt plus 3pt \everypar{\hangindent=\parindent}}

\gdef\refis#1{\item{#1.\ }}			

\gdef\journal#1, #2, #3, 1#4#5#6{		
    {\sl #1~}{\bf #2}, #3 (1#4#5#6)}		

\def\endreferences{\body}

\def\figurecaptions		
  {\endpage
   \beginparmode
   \head{Figure Captions}
}

\def\endpage			
  {\vfill\eject}

\def\endpaper			
  {\endmode\vfill\supereject}


\def\heading				
  {\vskip 0.5truein plus 0.1truein	
   \beginparmode \def\\{\par} \parskip=0pt \singlespace \raggedcenter}

\def\subheading				
  {\vskip 0.25truein plus 0.1truein	
   \beginlinemode \singlespace \parskip=0pt \def\\{\par}\raggedcenter}

\def\tag#1$${\eqno(#1)$$}

\def\align#1$${\eqalign{#1}$$}

\def\aligntag#1$${\gdef\tag##1\\{&(##1)\cr}\eqalignno{#1\\}$$
  \gdef\tag##1$${\eqno(##1)$$}}

\def\endaligntag{}

\def\overset #1\to#2{{\mathop{#2}\limits^{#1}}}
\def\underset#1\to#2{{\let\next=#1\mathpalette\undersetpalette#2}}
\def\undersetpalette#1#2{\vtop{\baselineskip0pt
\ialign{$\mathsurround=0pt #1\hfil##\hfil$\crcr#2\crcr\next\crcr}}}


\def\ref#1{Ref.~#1}			
\def\Ref#1{Ref.~#1}			
\def\[#1]{[\cite{#1}]}
\def\cite#1{{#1}}
\def\(#1){(\call{#1})}
\def\call#1{{#1}}
\def\taghead#1{}
\def\frac#1#2{{#1 \over #2}}

\def\12{{1\over2}}

\def\sla{\raise.15ex\hbox{$/$}\kern-.57em}
\def\leaderfill{\leaders\hbox to 1em{\hss.\hss}\hfill}
\def\twiddle{\lower.9ex\rlap{$\kern-.1em\scriptstyle\sim$}}
\def\bigtwiddle{\lower1.ex\rlap{$\sim$}}
\def\gtwid{\mathrel{\raise.3ex\hbox{$>$\kern-.75em\lower1ex\hbox{$\sim$}}}}
\def\ltwid{\mathrel{\raise.3ex\hbox{$<$\kern-.75em\lower1ex\hbox{$\sim$}}}}
\def\square{\kern1pt\vbox{\hrule height 1.2pt\hbox{\vrule width 1.2pt\hskip 3pt
   \vbox{\vskip 6pt}\hskip 3pt\vrule width 0.6pt}\hrule height 0.6pt}\kern1pt}
\def\tdot#1{\mathord{\mathop{#1}\limits^{\kern2pt\ldots}}}

\def\pmb#1{\setbox0=\hbox{#1}%
  \kern-.025em\copy0\kern-\wd0
  \kern  .05em\copy0\kern-\wd0
  \kern-.025em\raise.0433em\box0 }

\def\3he{{$^3${\rm He}}}

\def\]{\right]}
\def\[{\left[}

\def\>{\rangle}
\def\<{\langle}

\def\slD{\raise.15ex\hbox{$/$}\kern-.57em\hbox{$D$}}
\def\dsl{\raise.15ex\hbox{$/$}\kern-.57em\hbox{$\Delta$}}
\def\slp{{\raise.15ex\hbox{$/$}\kern-.57em\hbox{$\partial$}}}
\def\nsl{\raise.15ex\hbox{$/$}\kern-.57em\hbox{$\nabla$}}
\def\sla{\raise.15ex\hbox{$/$}\kern-.57em\hbox{$\rightarrow$}}
\def\slla{\raise.15ex\hbox{$/$}\kern-.57em\hbox{$\lambda$}}
\def\slb{\raise.15ex\hbox{$/$}\kern-.57em\hbox{$b$}}
\def\lnp{\raise.15ex\hbox{$/$}\kern-.57em\hbox{$p$}}
\def\lnk{\raise.15ex\hbox{$/$}\kern-.57em\hbox{$k$}}
\def\lnK{\raise.15ex\hbox{$/$}\kern-.57em\hbox{$K$}}
\def\lnq{\raise.15ex\hbox{$/$}\kern-.57em\hbox{$q$}}


\def\pmb#1{\setbox0=\hbox{$#1$}%
\kern-.025em\copy0\kern-\wd0
\kern.05em\copy0\kern-\wd0
\kern-.025em\raise.0433em\box0 }

\def\q2{{Q^2}}
\def\gtwid{\raise.3ex\hbox{$>$\kern-.75em\lower1ex\hbox{$\sim$}}}
\def\ltwid{\raise.3ex\hbox{$<$\kern-.75em\lower1ex\hbox{$\sim$}}}
\def\12{{1\over2}}
\def\part{\partial}

\def\low#1{\lower.5ex\hbox{${}_#1$}}

\def\psl{\raise.15ex\hbox{$/$}\kern-.57em\hbox{$\partial$}}
\def\partt{\raise.15ex\hbox{$\widetilde$}{\kern-.37em\hbox{$\partial$}}}

\def\abs{
         \vskip 3pt plus 0.3fill\beginparmode
         \doublespace ABSTRACT:\ }

\def\topppageno1{\global\footline={\hfil}\global\headline
={\ifnum\pageno<\firstpageno{\hfil}\else{\hss\twelverm --\ \folio
\ --\hss}\fi}}

\def\toppageno2{\global\footline={\hfil}\global\headline
={\ifnum\pageno<\firstpageno{\hfil}\else{\rightline{\hfill\hfill
\twelverm \ \folio
\ \hss}}\fi}}

\catcode`@=11
\newcount\r@fcount \r@fcount=0
\newcount\r@fcurr
\immediate\newwrite\reffile
\newif\ifr@ffile\r@ffilefalse
\def\w@rnwrite#1{\ifr@ffile\immediate\write\reffile{#1}\fi\message{#1}}

\def\writer@f#1>>{}
\def\referencefile{
  \r@ffiletrue\immediate\openout\reffile=\jobname.ref%
  \def\writer@f##1>>{\ifr@ffile\immediate\write\reffile%
    {\noexpand\refis{##1} = \csname r@fnum##1\endcsname = %
     \expandafter\expandafter\expandafter\strip@t\expandafter%
     \meaning\csname r@ftext\csname r@fnum##1\endcsname\endcsname}\fi}%
  \def\strip@t##1>>{}}

\def\citeall#1{\xdef#1##1{#1{\noexpand\cite{##1}}}}
\def\cite#1{\each@rg\citer@nge{#1}}	

\def\each@rg#1#2{{\let\thecsname=#1\expandafter\first@rg#2,\end,}}
\def\first@rg#1,{\thecsname{#1}\apply@rg}	
\def\apply@rg#1,{\ifx\end#1\let\next=\relax
\else,\thecsname{#1}\let\next=\apply@rg\fi\next}

\def\citer@nge#1{\citedor@nge#1-\end-}	
\def\citer@ngeat#1\end-{#1}
\def\citedor@nge#1-#2-{\ifx\end#2\r@featspace#1 
  \else\citel@@p{#1}{#2}\citer@ngeat\fi}	
\def\citel@@p#1#2{\ifnum#1>#2{\errmessage{Reference range #1-#2\space is bad.}%
    \errhelp{If you cite a series of references by the notation M-N, then M and
    N must be integers, and N must be greater than or equal to M.}}\else%
 {\count0=#1\count1=#2\advance\count1
by1\relax\expandafter\r@fcite\the\count0,%
  \loop\advance\count0 by1\relax
    \ifnum\count0<\count1,\expandafter\r@fcite\the\count0,%
  \repeat}\fi}

\def\r@featspace#1#2 {\r@fcite#1#2,}	
\def\r@fcite#1,{\ifuncit@d{#1}
    \newr@f{#1}%
    \expandafter\gdef\csname r@ftext\number\r@fcount\endcsname%
                     {\message{Reference #1 to be supplied.}%
                      \writer@f#1>>#1 to be supplied.\par}%
 \fi%
 \csname r@fnum#1\endcsname}
\def\ifuncit@d#1{\expandafter\ifx\csname r@fnum#1\endcsname\relax}%
\def\newr@f#1{\global\advance\r@fcount by1%
    \expandafter\xdef\csname r@fnum#1\endcsname{\number\r@fcount}}

\let\r@fis=\refis			
\def\refis#1#2#3\par{\ifuncit@d{#1}
   \newr@f{#1}%
   \w@rnwrite{Reference #1=\number\r@fcount\space is not cited up to now.}\fi%
  \expandafter\gdef\csname r@ftext\csname r@fnum#1\endcsname\endcsname%
  {\writer@f#1>>#2#3\par}}

\def\ignoreuncited{
   \def\refis##1##2##3\par{\ifuncit@d{##1}%
     \else\expandafter\gdef\csname r@ftext\csname
r@fnum##1\endcsname\endcsname%
     {\writer@f##1>>##2##3\par}\fi}}

\def\r@ferr{\endreferences\errmessage{I was expecting to see
\noexpand\endreferences before now;  I have inserted it here.}}
\let\r@ferences=\references
\def\references{\r@ferences\def\endmode{\r@ferr\par\endgroup}}

\let\endr@ferences=\endreferences
\def\endreferences{\r@fcurr=0
  {\loop\ifnum\r@fcurr<\r@fcount
    \advance\r@fcurr by 1\relax\expandafter\r@fis\expandafter{\number\r@fcurr}%
    \csname r@ftext\number\r@fcurr\endcsname%
  \repeat}\gdef\r@ferr{}\endr@ferences}


\let\r@fend=\endpaper\gdef\endpaper{\ifr@ffile
\immediate\write16{Cross References written on []\jobname.REF.}\fi\r@fend}

\catcode`@=12

\citeall\refto		
\citeall\ref		%
\citeall\Ref		%

\catcode`@=11
\newcount\tagnumber\tagnumber=0

\immediate\newwrite\eqnfile
\newif\if@qnfile\@qnfilefalse
\def\write@qn#1{}
\def\writenew@qn#1{}
\def\w@rnwrite#1{\write@qn{#1}\message{#1}}
\def\@rrwrite#1{\write@qn{#1}\errmessage{#1}}

\def\taghead#1{\gdef\t@ghead{#1}\global\tagnumber=0}
\def\t@ghead{}

\expandafter\def\csname @qnnum-3\endcsname
  {{\t@ghead\advance\tagnumber by -3\relax\number\tagnumber}}
\expandafter\def\csname @qnnum-2\endcsname
  {{\t@ghead\advance\tagnumber by -2\relax\number\tagnumber}}
\expandafter\def\csname @qnnum-1\endcsname
  {{\t@ghead\advance\tagnumber by -1\relax\number\tagnumber}}
\expandafter\def\csname @qnnum0\endcsname
  {\t@ghead\number\tagnumber}
\expandafter\def\csname @qnnum+1\endcsname
  {{\t@ghead\advance\tagnumber by 1\relax\number\tagnumber}}
\expandafter\def\csname @qnnum+2\endcsname
  {{\t@ghead\advance\tagnumber by 2\relax\number\tagnumber}}
\expandafter\def\csname @qnnum+3\endcsname
  {{\t@ghead\advance\tagnumber by 3\relax\number\tagnumber}}

\def\equationfile{%
  \@qnfiletrue\immediate\openout\eqnfile=\jobname.eqn%
  \def\write@qn##1{\if@qnfile\immediate\write\eqnfile{##1}\fi}
  \def\writenew@qn##1{\if@qnfile\immediate\write\eqnfile
    {\noexpand\tag{##1} = (\t@ghead\number\tagnumber)}\fi}
}

\def\callall#1{\xdef#1##1{#1{\noexpand\call{##1}}}}
\def\call#1{\each@rg\callr@nge{#1}}

\def\each@rg#1#2{{\let\thecsname=#1\expandafter\first@rg#2,\end,}}
\def\first@rg#1,{\thecsname{#1}\apply@rg}
\def\apply@rg#1,{\ifx\end#1\let\next=\relax%
\else,\thecsname{#1}\let\next=\apply@rg\fi\next}

\def\callr@nge#1{\calldor@nge#1-\end-}
\def\callr@ngeat#1\end-{#1}
\def\calldor@nge#1-#2-{\ifx\end#2\@qneatspace#1 %
  \else\calll@@p{#1}{#2}\callr@ngeat\fi}
\def\calll@@p#1#2{\ifnum#1>#2{\@rrwrite{Equation range #1-#2\space is bad.}
\errhelp{If you call a series of equations by the notation M-N, then M and
N must be integers, and N must be greater than or equal to M.}}\else%
 {\count0=#1\count1=#2\advance\count1
by1\relax\expandafter\@qncall\the\count0,%
  \loop\advance\count0 by1\relax%
    \ifnum\count0<\count1,\expandafter\@qncall\the\count0,%
  \repeat}\fi}

\def\@qneatspace#1#2 {\@qncall#1#2,}
\def\@qncall#1,{\ifunc@lled{#1}{\def\next{#1}\ifx\next\empty\else
  \w@rnwrite{Equation number \noexpand\(>>#1<<) has not been defined yet.}
  >>#1<<\fi}\else\csname @qnnum#1\endcsname\fi}

\let\eqnono=\eqno
\def\eqno(#1){\tag#1}
\def\tag#1$${\eqnono(\displayt@g#1 )$$}

\def\aligntag#1\endaligntag
  $${\gdef\tag##1\\{&(##1 )\cr}\eqalignno{#1\\}$$
  \gdef\tag##1$${\eqnono(\displayt@g##1 )$$}}

\def\eqalignno#1{\displ@y \tabskip\centering
  \halign to\displaywidth{\hfil$\displaystyle{##}$\tabskip\z@skip
    &$\displaystyle{{}##}$\hfil\tabskip\centering
    &\llap{$\displayt@gpar##$}\tabskip\z@skip\crcr
    #1\crcr}}

\def\displayt@gpar(#1){(\displayt@g#1 )}

\def\displayt@g#1 {\rm\ifunc@lled{#1}\global\advance\tagnumber by1
        {\def\next{#1}\ifx\next\empty\else\expandafter
        \xdef\csname @qnnum#1\endcsname{\t@ghead\number\tagnumber}\fi}%
  \writenew@qn{#1}\t@ghead\number\tagnumber\else
        {\edef\next{\t@ghead\number\tagnumber}%
        \expandafter\ifx\csname @qnnum#1\endcsname\next\else
        \w@rnwrite{Equation \noexpand\tag{#1} is a duplicate number.}\fi}%
  \csname @qnnum#1\endcsname\fi}

\def\ifunc@lled#1{\expandafter\ifx\csname @qnnum#1\endcsname\relax}

\let\@qnend=\end\gdef\end{\if@qnfile
\immediate\write16{Equation numbers written on []\jobname.EQN.}\fi\@qnend}

\catcode`@=12


\def\quark{(\bar s_L\gamma^\mu t^a d_L)\,\, (\bar u_L\gamma_\mu t^a u_L)}
\def\quarka{\bar u_L\gamma_\mu t^a u_L}
\def\quarkb{\bar s\gamma^\mu\gamma^5 t^a g\tilde G^a_{\alpha\mu}  d}

\def\for{\tilde f_+}
\rightline{NSF-ITP-92-82}
\rightline{TPI-MINN-92-26-T}
\rightline{May  1992}
\vskip.8in
\centerline{\bf Nonleptonic decays of K mesons revisited}
\bigskip
\centerline{B. Blok
}
\bigskip
\centerline{\sl Institute for Theoretical Physics}
\centerline{\sl University of California at Santa Barbara}
\centerline{\sl Santa Barbara, CA 93106 }
\centerline{\sl and}
\centerline{M. Shifman
}
\bigskip
\centerline{\sl  Theoretical Physics Institute}
\centerline{\sl University of Minnesota}
\centerline{\sl Minnesota, MN 55455}
\bigskip
\abs{ We estimate nonfactorizable 1/$N_c$ contributions in the $K\rightarrow
2\pi$ amplitudes using the approach proposed in our previous work. It is
demonstrated that for the conventional (nonpenguin) operators these
 contributions are close in magnitude to factorizable $1/N_c$ parts and have
the opposite sign. Thus, an approximate rule of discarding $1/N_c$ corrections
in $K\rightarrow 2\pi$ decays
 is theoretically confirmed. As a result we find an extra
suppression  of the matrix element of $O_4$ ($\Delta I=3/2$) and an extra
enhancement of the matrix element  of $O_1$ ($\Delta I=1/2$).
 The parameter $B$ describing
$K^0-\bar K^0$ mixing is also discussed.}
\endpage
\par 1. Twenty years after creation of QCD the nonleptonic decays of K mesons
continue to attract attention. Understanding of the underlying dynamics
$-$ which,
unfortunately, is still incomplete $-$
 is important for other applications, like
the study of CP violation and constraints on the t-quark mass and new physics.
The basic elements of the modern theory of nonleptonic K-decays were laid in
\ref{1,2}, and are described in the standard textbooks (see e.g. \ref{3}). We
will refer to \ref{2} as SVZ, and will consistently use the notations accepted
in \ref{2, 3}.
\par There is an ongoing controversy in the literature concerning
 the coefficient
$c_5$ and the role of the penguin operators in general. Leaving this aspect
 aside we will concentrate here on the nonpenguin operators $O_1$ and $O_4$
and discuss the rule of discarding $1/N_c$ corrections in
 these operators.\refto{4}
It has been empirically observed \refto{4} (for a pedagogical discussion
see \ref{5}) that supplementing the SVZ factorization in $O_1$ and $O_4$ by
a prescription to disregard all $1/N_c$
terms altogether enhances the $K\rightarrow 2\pi$ matrix element of $O_1$
($\Delta I=1/2$) and suppresses the $K\rightarrow 2\pi$ matrix element of $O_4$
($\Delta I=3/2)$. A few attempts to find a dynamical explanation for this rule
have been undertaken. In particular, in \ref{6,7,8} it is suggested to describe
the nonfactorizable parts in terms of hadronic interactions within the chiral
perturbation theory. Chiral loops (with one extra vertex) generate $1/N_c$
corrections which tend to cancel $1/N_c$ terms appearing at the tree level.
\refto{6,7,8}
\par In the present paper we extend the approach of \ref{9} to K mesons
and suggest a framework allowing one to estimate nonfactorizable parts
essentially within QCD, with very mild  assumptions that can be
 rather reliably controlled theoretically. In general terms our mechanism can
be
pictorially described as follows: non-factorizable parts in a certain
 approximation are due to one soft gluon exchanged between two quark brackets
forming $O_1$ and $O_4$ (see below). Formally, the result reduces to the
$K\pi$ matrix element of the operator
$$\bar s \gamma^\mu(1+\gamma^5)\tilde G_{
\alpha\mu}d.\eqno (1)$$
A key role played by the operators of this type (with different quark flavors)
in the nonfactorizable $1/N_c$ amplitudes has been revealed in \ref{9}.
In the heavy quark transition the matrix element of
$\bar Q_1 \gamma^\mu (1+\gamma^5)\tilde G_{
\alpha\mu}Q_2$ is equivalent to a correlation of the heavy  quark spin
with the chromomagnetic field, and is proportional to the (experimentally
measured) spin splitting between $B^*$ and $B$. Clearly, in the kaon decays
one can not use the same techniques to fix the $K\pi$ matrix element of the
operator \(1) . Fortunately, it is reliably known from other sources
(\ref{10,11}). Calculations presented below rely on some
 ideas and results obtained from  the QCD sum rules
method.
\refto{12} We do not use, however, all heavy  machinery of this method. The
elements involved are rather transparent.
\par The most remarkable finding is that our QCD-based calculations confirm the
empirical rule \refto{4} of discarding  $1/N_c$ corrections in the kaon decays,
just in the same way as it has been demonstrated recently for the heavy meson
decays \refto{9}. As it was noticed previously
\refto{4,5}, the implications of this rule
are very favorable: the $O_1$ contribution ($\Delta I=1/2$) is enhanced by a
factor $ 3/2$ compared to the original SVZ estimates
(done with naive factorization)
 while that of $O_4$ $(\Delta I=3/2)$ is suppressed
by a factor $ 3/4$.
\par 2. Let us recall now the SVZ  procedure of calculating
the matrix elements of the four-fermion operators in $K\rightarrow 2\pi$
decay. Consider, for definitness, the operator $O_4$:
$$O_4=(\bar s_L\gamma_\mu d_L)(\bar u_L\gamma_\mu u_L)+
(\bar s_L\gamma_\mu u_L)(\bar u_L\gamma_\mu d_L)-
(\bar s_L\gamma_\mu d_L)(
\bar d_L\gamma_\mu d_L).\eqno (2)$$
The relevant diagrams are the spectator graphs of Figs. 1 a$-$c
 (the contribution
of Fig. 1d vanishes, see below).
\par For the spectator mechanism one can show \refto{2} that Figs. 1 a$-$c give
one and the same contribution which is equal to the matrix element of the
operator
$$O_4\vert_{Fig.1 a}\rightarrow
(1+{1\over N_c})(\bar s_L\gamma^\mu d_L)(\bar u_L\gamma_\mu u_L)+2
(\bar s_L\gamma^\mu t^ad_L)(
\bar u_L\gamma_\mu t^au_L).\eqno (3)$$
Here $t^a$ are color matrices, $Tr(t^at^b)={1\over 2}\delta^{ab}$.
The naive factorization procedure prescribes to saturate the four fermion
operator \(3) by the vacuum  intermediate state, e.g.:
$$<\pi^+\pi^0\vert
 (\bar s_L\gamma^\mu d_L)(\bar u_L\gamma_\mu u_L)\vert K^+>_{Fig. 1a}=
<\pi^0\vert \bar u_L\gamma^\mu u_L\vert 0><\pi^+\vert \bar s_L\gamma_\mu d_L
\vert K^+>.\eqno (4)$$
The second operator in the
 r.h.s. of eq. \(3) is totally disregarded in this procedure
since no color singlet intermediate state is possible. Then one obtains
$$<\pi^+\pi^0\vert O_4\vert K^+>_{\rm f}
=3(1+{1\over N_c}){if_\pi\over 4\sqrt{2}}
(f_+m^2_K).\eqno (5)$$
Here the subscript "f" stands for factorizable and the factor 3 comes from
adding together Figs. 1a$-$1c.
The numerical constants in this expression are equal to
$f_\pi=133\quad {\rm MeV}$, $f_+=1$ at zero momentum transfer.
  We make use of the fact that
$$\eqalign{
&<\pi^0\vert u_L\gamma_\mu u_L\vert 0>=-{if_\pi\over 2\sqrt{2}}q_\mu ,\cr
&<\pi^+\vert \bar s_L\gamma_\mu d_L\vert K^+>=-{1\over 2}(p_K+p_\pi)_{\mu}f_+,
\cr
}\eqno (6)$$
In eqs. \(5), \(6) we actually neglect some small and irrelevant terms
(e. g. O$(m^2_\pi)$ in eq. \(5), $f_-$ in eq. \(6) , etc. ).
To the leading order in $1/N_c$ the procedure described above and eq. \(5) are
exact \refto{13a} in the multicolor chromodynamics \refto{13}
(for a review see \ref{14a}). The
$O(1/N_c)$ term in eq. \(5) is not justified by the $1/N_c$ counting,
however, and $-$ as we
 will see shortly $-$ is cancelled by the nonfactorizable
second term in eq. \(3) . Before proceeding to the analysis of
 the nonfactorizable parts let us mention that for the $\Delta I=1/2$
transition $K^0\rightarrow \pi^+\pi^-$ (Fig. 2)
$$\eqalign{&
<\pi^+\pi^-\vert O_1\vert K^0>_{\rm f}=({1\over N_c}-1)
<\pi^+\vert \bar u_L\gamma^\mu d_L\vert 0><\pi^-\vert \bar s_L\gamma_\mu u_L
\vert K^0>\cr
&=({1\over N_c}-1){if_\pi\over 4}m^2_K,\cr}\eqno (8)$$
while
 $$<\pi^+\pi^-\vert O_1\vert K^0>_{\rm n.f.}=<\pi^+\pi^-\vert
2(\bar s_L\gamma^\mu t^au_L)(\bar u_L\gamma_\mu t^a d_L)\vert K^0>.\eqno (9)$$
The subscript "n.f." is used to mark the nonfactorizable parts.
\par Our task now is to estimate the nonfactorizable parts $-$ the second
term in the r.h.s. of eq. \(3) and r.h.s. of eq. \(9) .
 To this end we use a simple and transparent method
of \ref{9}.  Referring the reader to that paper for a detailed description we
just sketch here  basic stages. The starting idea is to split the calculation
in two parts. At first, instead of the $K\rightarrow 2\pi$ matrix
elements  of operators $O_{1,4}$
we consider the correlation function of $O_{1,4}$ with an appropriately chosen
axial current $A^\beta$. The current is to be chosen in such a way as to
produce
a required pion from the vacuum. For instance, if $\pi^0$ is to be treated in
 this way, we introduce
$$ A^\beta ={1\over \sqrt{2}}(\bar u\gamma^\beta\gamma^5 u-\bar d\gamma^\beta
\gamma^5 d).\eqno (10)$$
The correlation function of $O_4$ and $ A^\beta$ is sandwiched between
$K^+$ and $\pi^+$ states,
$${\cal A}^\beta=<\pi^+\vert \int d^4x e^{iqx}<\pi^+\vert T\{O_4(0) A^\beta (x)
\}\vert K^+>.\eqno (11)$$
The momentum q is an auxiliary momentum ( it enters the graph at the point
$x$ and leaves it at $0$)  assumed to be Euclidean. If $-q^2=Q^2\sim 1$
GeV$^2$, the distance between the points $x$ and $0$ is small and we can
 consistently use the standard operator product expansion (OPE) to calculate
the correlation function \(11) by expanding the r.h.s. of eq. \(11) in
inverse powers of $Q^2$. Since we are interested in the nonfactorizable parts
only, we will keep only the the second term in eq. \(3) . It is implied that
the contribution of Fig. 1a is analysed. We will suppress the corresponding
 subscript. For graphs Figs. 1b and 1c the results are the same, while
the graphs of Fig. 1d do not contribute. ( One can easily check that the
vanishing of the factorizable part of Fig. 1d persists for the nonfactorizable
part as well in our approximation).
\par Technically, the operator product expansion
for $<2\quark , A^\beta>$ is the same as for the corresponding correlator in
\ref{9}.
The only difference is an obvious substitution of the heavy quark bracket by
$(\bar s_L\gamma^\mu t^a d_L)$.
 One has to fuse $A^\beta$ with the bracket
$\quarka$. The leading operator emerging in this fusion is due to the soft
gluon
emission (Fig. 3):
$$\eqalign{
{\cal A}^\beta & =-{i\over 4\pi^2}{1\over 4\sqrt{2}}<\pi^+\vert
\quarkb\vert K^+> {q^\alpha q^\beta\over q^2}\cr
&+{\rm terms}\quad {\rm suppressed}\quad {\rm by}\quad {\rm powers}
\quad {\rm of }\quad Q^{-2}.\cr}\eqno (12)$$
 \par The expression given in eq. \(12) contains a pion-like pole, and it
is
 tempting to interpret it as exclusively due to pion. Clearly, this is not
 the case since we can not trust eq. \(12) at $Q^2\rightarrow 0$, where
higher
order operators come into play and blow up. Eq. \(12) is valid only
 at sufficiently large $Q^2$ ($Q^2\sim 1$ GeV$^2$).
 Nevertheless, from what we know about
the
axial current $A^{\beta}$, we can
 say that pion constitutes a sizable fraction of all intermediate states $-$
there is a pion dominance in $A^{\beta}$ just in the same way as
there is the
$\rho-$meson dominance in the vector current. This observation will
allow us
to extract the pion contribution from eq. \(12).

\par Saturating eq. \(12) by  $\pi^0$
 and amputating the pion leg
(i.e. dividing by $-f_\pi q^\beta/q^2$) we arrive at the following expression
$$<\pi^+\pi^0\vert O_4\vert K^+>_{\rm n.f.}
={i\over 4\pi^2 f_\pi}{1\over 4\sqrt{2}}
q^\alpha <\pi^+\vert\quarkb\vert K^+>.\eqno (13)$$
(Warning: this is the contribution corresponding to Fig. 1a; in the total
 amplitude one has to include Figs. 1b,c).
\par Surprising though
 it is, the matrix element in the r.h.s. of eq. \(13) which
plays a crucial role in our calculations is actually known for a long time
\refto{10,11}. On general grounds one can write that
$$<\pi^+\vert\quarkb\vert
 K^+>=\tilde f_+(p_K+p_\pi)_\alpha +\tilde f_- (p_K-p_\pi)
_\alpha .\eqno (14)$$
Here $\tilde f_{\pm}$ are analogs of the conventional formfactors
$f_{\pm}$ that arise in the amplitudes of the semileptonic decays
 $ <\pi^+\vert\bar s_L\gamma^\mu d_L\vert K^+>$. Generally speaking, $\tilde
f_{\pm }$ are functions of the momentum transfer squared, just like the
conventional $f_{\pm}$. In the problem at hand the momentum transfer squared
$q^2=m^2_\pi  $ and
 we approximate $\tilde f_{\pm}$  by constants at zero momentum
transfer.
 Moreover, the ratio ${\tilde  f_-\over
\tilde f_+}$ vanishes in the
 SU(3)$_{\rm fl}$ flavor limit, when $m_s\rightarrow
 m_u$. The easiest way to see that this is indeed the case
 is to reduce (with the
help of the standard PCAC) (i) pion, leaving $K^+$ intact, or (ii) kaon,
 leaving $\pi^+$ intact. We immediately get that in the SU(3)$_{\rm fl}$
symmetry limit
$\tilde f_+-\tilde f_-=\tilde f_+ +\tilde f_-$. Hence, $\tilde f_-$ can be
 neglected. We finally arrive at the following formula for
$\for$:
$$\eqalign{\for (p_K)_\alpha &={i\over f_\pi}<0\vert\bar s\gamma^\mu
 t^ag\tilde G^a_{\alpha\mu} d
\vert K^+>\cr
           &=m_1^2(p_K)_\alpha .\cr}\eqno (15)$$
A numerical constant $m_1^2$ has dimension GeV$^2$. It parametrizes the kaon
coupling to the current
 $$\tilde a_\alpha =\bar s\gamma^\mu
 t^ag\tilde G^a_{\alpha\mu} d\eqno (16)$$
This constant was first introduced in \ref{10}.
(More exactly, its non-strange
 analog. They, however, are equal in the chiral limit
 we consider here). From \ref{10}
we have:
$$<0\vert \tilde a _\alpha\vert K^+>=if_\pi m_1^2(p_K)_\alpha .\eqno (17)$$
The constant $m_1^2$ is positive. (Warning: the current \(16) and the
 corresponding current in
 \ref{10} differ in signs. The parameter $m_1^2$ used here was denoted by
$\delta^2$ in \ref{10}.)
 The constant $m_1^2$ can be determined using
QCD sum rules and OPE methods. \refto{10,11}
 Another way to determine it is from the
study of higher twist corrections to the amplitude $\pi^0\rightarrow
\gamma^*\gamma^*$ ($\gamma^*$ stands for an off-shell photon) using the
 Brodsky-Lepage approach to exclusive processes. \refto{101}
Both estimates give
$$m_1^2\simeq 0.2 \quad{\rm GeV}^2.\eqno (17a)$$
The positivity of $m_1^2$, that will play a crucial role in the cancellation
of $1/N_c$ corrections is rather natural. Indeed, by virtue of equations of
 motion
$$\tilde a_\alpha=\bar s\gamma_\alpha\gamma^5(-{\cal P})^2d.\eqno (19)$$
Here ${\cal P}$
is the  momentum operator: ${\cal P_\alpha}=i{\cal D_\alpha}$. One can
expect that for the bound state $(-{\cal P})^2\rightarrow m_1^2$.
\par Combining eqs. \(15) and \(17) we conclude that
$\tilde f_+=-m_1^2$ and
$$<\pi^+\vert \quarkb\vert K^+>\simeq -m_1^2(p_K+p_\pi)_\alpha .\eqno (20)$$
\par We are now in position to determine the nonfactorizable part of the
$\Delta I$=3/2 amplitude. From eqs. \(13) and \(20) it follows immediately
$$<\pi^+\pi^0\vert O_4\vert K^+>_{\rm n.f.}=-{im_1^2\over 4\pi^2f_\pi}
{3\over 4\sqrt{2}}m_K^2.\eqno (21)$$
Here we restored the factor 3 that reflects the additive and equal
contributions
of Figs. 1a$-$c.
\par Let us now compare eq. \(21) with the factorizable
 part of the same amplitude suppressed by $1/N_c$
(eq. \(5)). The ratio
$$r={<\pi^+\pi^0\vert O_4\vert K^+>_{\rm n.f.}
\over <\pi^+\pi^0\vert O_4\vert K^+>_{{\rm f},1/N}}
=-{N_cm^2_1\over 4\pi^2f_\pi^2}.
\eqno (22)$$
In the denominator of eq. \(22)  we keep only ${1\over N_c}$ part of
the factorizable
amplitude by definition.
As was expected $r$ is $O(N_c^0)$, and as it was announced before, the ratio
$r$ is very close to $-1$.
The sources of uncertainties in our determination of $r$ are as follows:
\par (i) the approximation of the pion dominance (possible higher state
 contamination);
\par (ii) corrections due
 to the deviations from the chiral limit in calculating
 matrix elements like $m_1^2$;
\par (iii) corrections due to the operators with dimension $>2$ and deviations
from the chiral limit in the operator product expansion that led to eq. \(12).
\par  All these  uncertainties are expected to be modest.
We will discuss them later.
\par 4. Let us proceed now to the $\Delta I=1/2$ transition. We merely
repeat
 the procedure described above for the amplitude $K^0\rightarrow \pi^+\pi^-
$. We limit ourselves to the contribution of
the operator $O_1$  since other
contributions (due to $O_2$ and $O_3$)  are negligible. It
is straightforward to check that in this case the nonfactorizable part
of the amplitude reduces to the matrix element of the  operator
$$\tilde O =2(\bar s_L\gamma^\mu t^a u_L)(\bar u_L\gamma_\mu t^a d_L).
\eqno (22a)$$
We see that the nonfactorizable part of the amplitude is equal to the matrix
element of the same type operator as for the
 $\Delta I =3/2$ transition sandwiched between
the states $K^0$ and $\pi^+\pi^-$.
This only changes  the auxiliary current $A^\beta$ which  has now the form
$A^\beta =\bar d\gamma^\beta\gamma^5 u$.
 Repeating the procedure described above
for the latter transition we arrive at the matrix element
$<\pi^-\vert \bar s\gamma^{\mu}\gamma^5t^ag\tilde G^a_{\alpha\mu} u\vert K^0>
$ and get
$$<\pi^+\pi^-\vert O_1\vert K^0>_{\rm n.f.}=-{im_1^2\over 4\pi^2f_\pi}
{1\over 4}m^2_K.\eqno (219)$$
This expression must be compared with the $1/N_c$ suppressed term in the
factorized amplitude, eq. \(8) .
We get for the corresponding ratio
$$r'={<\pi^+\pi^-\vert O_1\vert K^0>_{\rm n.f.}\over
<\pi^+\pi^-\vert O_1\vert K^0>_{{\rm f},1/N}}=r\sim -1.\eqno (161)$$
Note that the ratios of the $1/N_c$ factorizable and nonfactorizable amplitudes
are the same in both decays.
\par We have determined the
"normal" (nonpenguin) part of the $K^0\rightarrow\pi^-\pi^+$
amplitude due to the operator $O_1$. However there still exists uncertainty in
 the full amplitude of this decay. The reason is that, as it
was mentioned above, the major part of this amplitude is believed to
originate from
the penguin operators $O_5, O_6$. These operators form a part of the effective
weak Hamiltonian
$$H_{\rm weak}=c_5O_5+c_6O_6.\eqno (171)$$
The matrix elements of the penguin operators are proportional to
 $(1-{1\over N^2_c})$, i.e. the $1/N_c$ contribution is absent
at all (see e.g. \ref{5} ). The calculation
of nonfactorizable terms of order $1/N^2_c$ is beyond the accuracy of our
approach. First, it requires the study of the contribution of higher
 dimensional operators in the operator product expansion leading to eq. \(12).
Second, there can be $1/N^2_c$ corrections to the amplitudes that are nonzero
at $N_c\rightarrow\infty$. Hence
at the moment we have nothing to add to the discussion
about the  penguin contributions.

\par 5. Let
 us now briefly discuss the phenomenological implications of our results
for $K\rightarrow\pi\pi $  decays.
 First, we have found that the total $1/N_c$
part of the $K^+\rightarrow\pi^+\pi^0$ amplitude  is close to
 zero due to the cancellation
between factorizable and nonfactorizable terms in it. This result justifies
the use of the rule of discarding $1/N_c$ corrections in this decay \refto{4}.
The cancellation of $1/N_c$ terms in the  $K^+\rightarrow\pi^+\pi^0$
amplitude
completely eliminates the residual discrepancy between the
theory and experiment for this decay.
\par For the decay $K^0\rightarrow\pi^+\pi^-$ the cancellation of $1/N_c$ terms
in the matrix elements of $O_1$ also
justifies the application of $1/N_c$ rule \refto{4} to these decays. The
contribution
 of the operator $O_1$ is enhanced by a factor $\sim 1.5$, as it was
pointed out already in \ref{4,6,7,8}.
 It is certainly a welcome development, even
 though the contribution of this operator is responsible only for a
non-dominant
part of the amplitude of this decay. Nevertheless, the value of the total
amplitude increases, diminishing the disagreement between the theory and the
 experiment.
\par Finally, two minor comments.
 First, notice that if we assume
that the coefficients $c_{5,6}$  have the values given in \ref{2} and add
 the enhancement of the contribution of $O_1$  discussed above, we
get for the  $K^0\rightarrow\pi^+\pi^-$  amplitude
$${\cal A}(K^0\rightarrow \pi^+\pi^-)\sim 1.08 G_Ff_\pi m^2_K.\eqno (181)$$
This value is in good agreement with experiment, giving the explanation of
$\Delta I=1/2$ rule. However, the values of $c_{5,6}$ are
rather sensitive to details of computation and continue to be a subject of
 controversy (see e.g. \ref{5, 102} for
 a recent discussion). We will not touch
this issue here.
\par Second, note that unlike  the empiric approach of \ref{4} we can
now see when the rule of discarding $1/N_c$ corrections can be applied and
when it can not. This is of course due to its dynamical origin.

\par 6. Let us now consider
 the $K-\bar K^0$ mixing arising due to the box diagram.
 The so called B-parameter for $K-\bar K^0$ mixing is defined as follows:
$$<K^0\vert \bar s\Gamma^\alpha d\bar d\Gamma_\alpha s\vert \bar K^0>={8\over
3}
Bm^2_Kf^2_K.\eqno (191)$$
This parameter determines the mass difference of $K_L$ and $K_S$
(see e.g. \ref{102,103} for recent reviews). We can use the same method as was
exploited above for the
 $K\rightarrow \pi\pi $ decays to study the amplitude \(191).
The use of naive factorization gives $B=(1+1/N_c){3\over 4}$. \refto{104}
However saturating by the vacuum intermediate state,
 we loose the contribution of the operator
$$O_B=2\bar s\gamma^\mu\gamma^5 t^ad\bar s\gamma_\mu\gamma^5
 t^ad\eqno (201)$$
We can use our
method
 to find the nonfactorizable part of this amplitude.
The only difference is slightly unusual kinematics (see Fig. 4).
The auxiliary momentum flows through the weak Hamiltonian and the auxiliary
current $A^\beta$.
This  can decrease the accuracy of our results.
\par Repeating the procedure
 above we can find that the nonfactorizable part of B
 has the form
$${iq^\alpha\over 4\pi^2f_K}2<0\vert \bar s\gamma^\mu
 \tilde gG^a_{\alpha\mu}t^a d
\vert K>=-{2m^2_Km^2_1\over 4\pi^2}.\eqno (211)$$
The constant $m_1^2$ in this equation was defined in eq. \(15). The
contribution
of the $1/N_c$ suppressed factorizable part of the amplitude in eq. \(191)
 is ${2\over N_c}m^2_Kf^2_K$. We get for the ratio of the
$1/N_c$ factorizable and nonfactorizable parts
$$r=-{N_cm_1^2\over 4\pi^2f^2_K}\sim -0.7,\eqno (221)$$
i.e. the same result as in eq. \(22), with $f_\pi$ substituted by $f_K$.
 Clearly, with our accuracy we can not distinguish between $f_K$ and $f_\pi$
while calculating the matrix elements. The difference between $f^{-2}_K$ and
$f^{-2}_\pi$ measures the uncertainty due to the deviations from the chiral
limit while calculating matrix elements. Sticking to  eq. \(221) literally
we get that
$$B\simeq 0.83.\eqno (231)$$
Notice that this result for B virtually coincides with the one obtained by
Reinders and Yazaki \refto{115} using QCD sum rules approach ( they obtained
$B\sim 0.84$). On the other hand
we can not exclude that $r\simeq -1$ and $B\simeq 0.75$.
 The prediction of the
chiral perturbation theory is $B=0.66\pm 0.1$ \refto{105},
although one should be very  cautious
since the first loop changes the tree result by a factor of 2:
 chiral
 perturbation theory  does not have high accuracy.
\par 7. Summarizing, we can say the following.
We have found that nonfactorizable amplitudes  arising due to  color
exchange
between different brackets in the effective weak Hamiltonian
 play an important role in the K-meson decays.
 The approximate rule of discarding
$1/N_c$ corrections emerges dynamically. It works pretty well in the $K^{\pm}
\rightarrow \pi^{\pm}\pi^0$ channel where it leads to a good agreement
between the theory and the experiment. In $K^0\rightarrow\pi\pi$ decays
it enhances the nonpenguin contributions to these decays by a factor
of $\sim 1.5$, decreasing the gap between the theory
and experiment. The nonfactorizable amplitudes also
contribute significantly to the B-parameter, giving $B\sim 0.83$ to $0.75$
, in rough
agreement with the $1/N_c$ rule that gives $B= 3/4$. Our derivation is model
independent and goes within QCD.
\par Quite remarkably, it seems, from the results of
this paper and \ref{9}, there is a simple mechanism that is
 responsible for the
observable amplitudes
in all weak  hadronic nonleptonic decays, including the decays of both
 heavy and strange mesons. This is the exchange of one soft gluon
between color brackets in the operators in the effective Hamiltonian
of the type \(9) .
In all channels considered so far this mechanism leads
 to the emergence of an approximate rule
 of discarding $1/N_c$ corrections.
 Note however, that due to its dynamical nature, the rule is
only approximate, and its validity in each particular decay must be checked
separately; the degree of compensation of factorizable and nonfactorizable
amplitudes can be different in different channels. We are now able
to find, at least approximately, the degree of this compensation directly
within QCD framework.
\par Finally, let us discuss the accuracy of our results and whether it can be
improved. The main loopholes in our analysis are the following:
\par (i) We neglected the contribution of higher resonances in
 extracting the pion (kaon) contributions and passing from eq. \(12) to
eq. \(13) . The accuracy of this approximation is discussed in
\ref{9}.
\par (ii) We use chiral limit while calculating the matrix elements that
are used in the nonfactorizable amplitudes in $K\rightarrow 2\pi$ and
$K^0-\bar K^0$ transitions. This means that we can not really distinguish
between the values of the parameter $m_1^2$ for K and $\pi-$ mesons and between
$f_K$ and $f_\pi$ while using eqs. \(15) and \(17a) or eq. \(221). This leads
to uncertainty of order $f_K/f_\pi$.
\par (iii) We leave only the leading operator in the OPE expansion that leads
to eq. \(12). We neglect the corrections due to higher dimensional operators,
and also due to nonzero  s-quark mass  and different values of the quark
 condensates for the strange and nonstrange quarks. We expect the former
 corrections
to be suppressed by the parameter $\lambda\sim {\mu^2\over Q^2}$, where
$\mu$ is a typical quark momentum inside hadrons, $\mu\sim 0.3$ to $0.4$
GeV, while
$Q^2\sim $ 1 GeV$^2$. Thus we obtain $\lambda\sim 0.1$ to $0.2$,
 leading to a parametric
smallness of higher order corrections.
\par Being conservative one can say that our results are a solid indication on
the desired tendency. The signs are definitely fixed and the absolute
values of nonfactorizable parts are probably fixed up to a factor $\sim 1.5$.
\par Finally, note, that it is quite
  possible to
take into account
the nonzero
 strange quark mass and higher dimensional
operators in OPE \(12). We can  use a
 QCD sum rules  type expansion
 in order to
write a sum rule for weak amplitude. Our eq. \(13) will be the first term of
such expansion. In this way one can in principle improve the accuracy of our
approach. The price will be, however, that we shall need additional
matrix element beyond, say, $m_1^2$, and they
 are not known
at present.
\endpage
\references
%
\refis{13a} A. A. Migdal, 1982, unpublished.

\refis{13} G.'t Hooft, Nucl. Phys., B72 (1974) 461;\hfill\break
          E. Witten, Nucl. Phys., B149 (1979) 285.


\refis{4} A. J. Buras, J.-M. Gerard and R. Ruckl, Nucl. Phys., B268
(1986) 16.

\refis{5} M. Shifman, Int. Journ. Mod. Phys., A3 (1988) 2769.

\refis{2} M. Shifman, A. Vainshtein and V. Zakharov,
           Nucl. Phys., B120 (1977)316; ZheTF, 72 (1977) 1275 [
           JETP, 45 (1977) 670].

\refis{10}  V. A. Novikov, M. A. Shifman, A. I. Vainshtein and V. I.
Zakharov and M. B. Voloshin, Nucl. Phys., B237 (1984) 525.

\refis{12} M. Shifman, A. Vainshtein and V. Zakharov, Nucl. Phys.,
B147
 (1979) 385, 448.


\refis{8} A. J. Buras and J.-M. Gerard, Nucl. Phys., B264 (1986) 371.

\refis{1} G. Altarelli and L. Maiani, Phys. Lett., B52, 351
(1974);\hfill\break
           M. K. Gaillard and B. W. Lee, Phys. Rev. Lett., 33 (1974) 108.

\refis{3} L. Okun, Leptons and Quarks, North Holland , 1982.

\refis{6} W. Bardeen,  A. J. Buras and J.M. Gerard, Nucl. Phys., B293(1987)
787.

\refis{7} W. Bardeen,  A. J. Buras and J.M. Gerard, Phys.Lett., 192B (1987)
          138.

\refis{105} W. Bardeen, A. J. Buras and J.M. Gerard, Phys.Lett.,
            B211 (1987) 343.

\refis{9} B. Blok and M. Shifman, "Nonfactorizable Amplitudes
              in Weak Nonleptonic Decays of Heavy Mesons",
               preprint NSF-ITP-92-76 [Submitted to Nucl. Phys. B].

\refis{11} V. L. Chernyak and A. I. Zhitnitsky, Phys. Rep., 112 (1984) 174.

\refis{102} H.-Y. Cheng, Int. Journ. of Mod. Phys., A4 (1989) 495.

\refis{103} E. A. Pachos and U. Turke, Phys. Rep., 178 (1989) 146.

\refis{115} L. J. Reinders and S. Yazaki, Nucl. Phys., B288 (1987) 789.

\refis{101} G. P. Lepage and S. J. Brodsky, Phys. Lett., B87 (1979); Phys. Rev.
             D22 (1980) 2157.

\refis{104} M. K. Gaillard and B. W. Lee, Phys. Rev., D10 (1974) 897.

\refis{14a} M. Shifman, in "Vacuum Structure and QCD sum rules", ed. M.
Shifman,
            North Holland, 1992.

\endreferences

\endpage

\centerline{\bf Figure Captions}
\bigskip
{\bf Fig.1:} Quark diagrams for $K^{+}\rightarrow \pi^+\pi^0$ decay.

{\bf Fig.2:} Quark diagram for $O_1$ contribution in
 $K^0\rightarrow \pi^+\pi^-$ decay.

{\bf Fig.3:} Color exchange amplitudes in the weak nonleptonic decays.

{\bf Fig.4:} Calculation of the nonfactorizable part of $K^0-\bar K^0$
             amplitude. The close circle

 denotes four-fermion operator
              $O_B$ (eq. \(201)).

\endpage
\endpaper
\end\bye